\documentclass[12pt]{article}
\usepackage{color}
\usepackage{latexsym}
\usepackage{epsfig,amssymb,euscript, mathrsfs}
\usepackage{amsmath}
\usepackage{graphicx}
\newsavebox{\ns}
\newsavebox{\dbrane}
\newsavebox{\dbshort}

\def\be{\begin{equation}}
\def\ee{\end{equation}}
\def\bea{\begin{eqnarray}}
\def\eea{\end{eqnarray}}

\newcommand{\nn}{\nonumber}

\newcommand\R{\mathbb{R}}
\newcommand\Z{\mathbb{Z}}

\newcommand\C{\mathbb{C}}

\newcommand\dd{\mathrm{d}}
\newcommand{\de}{\partial}

\newcommand{\ii}{\mathrm{i}}

\newcommand{\ex}{\mathrm{e}}
\newcommand{\vol}{\mathrm{vol}}
\newcommand{\Vol}{\mathrm{Vol}}
\newcommand{\Tr}{\mathrm{Tr}}

\newcommand{\ttau}{\sigma}
\newcommand{\anglepsi}{\varsigma}

\newcommand\e{\epsilon}

\newcommand{\xmax}{{x_\mathrm{max}}}
\newcommand{\xmin}{{x_\mathrm{min}}}

\newcommand\bz{\bar z}
\newlength{\sswidth}

\newcommand{\ppsi}{\vartheta}
\newcommand{\sign}{\mathrm{sign}}
\newcommand{\mfS}{\mathcal{S}}

\numberwithin{equation}{section}       

\begin{document}

\begin{titlepage}

\begin{center}

\today

\vskip 2.3 cm 

{\Large \bf  Wilson loops on three-manifolds and }

\vskip 0.5cm

{\Large \bf their M2-brane duals}

\vskip 2 cm

{Daniel Farquet and James Sparks\\}

\vskip 1cm

\textit{\it Mathematical Institute, University of Oxford,\\ 
\it Radcliffe Observatory Quarter, Woodstock Road, Oxford OX2 6GG, UK\\}

\end{center}

\vskip 2 cm

\begin{abstract}
\noindent We compute the large $N$ limit of Wilson loop expectation values for a broad 
class of $\mathcal{N}=2$ supersymmetric gauge theories defined on a general class 
of background three-manifolds $M_3$, diffeomorphic to $S^3$. We find a simple closed 
formula which depends on the background geometry only through a certain supersymmetric 
Killing vector field. The supergravity dual of such a Wilson loop is  an M2-brane
wrapping the M-theory circle, together with a complex curve $\Sigma_2$ in a self-dual 
Einstein manifold $M_4$, whose conformal boundary is $M_3$. We show that the regularized 
action of this M2-brane also depends only on the supersymmetric Killing vector, precisely 
reproducing the large $N$ field theory computation.
 
\end{abstract}

\end{titlepage}

\pagestyle{plain}
\setcounter{page}{1}
\newcounter{bean}
\baselineskip18pt

\tableofcontents

\section{Introduction and summary}

There has recently been considerable interest in defining and studying supersymmetric gauge theories on compact manifolds. 
This stems from the fact that certain observables may be computed exactly in such quantum field theories using localization. 
The first examples of such computations in the literature typically studied round sphere backgrounds, but 
more generally the observables also depend on the choice of background geometry, leading to a richer structure. 
Such exact computations may be used to test and explore non-perturbative dualities, and the focus of this 
paper will be the gauge/gravity duality.

In \cite{Alday:2013lba} the partition function $Z$ of three-dimensional $\mathcal{N}=2$ supersymmetric gauge theories 
on a general class of background three-manifold geometries $M_3$ was computed exactly. In particular 
$Z$ was shown to depend on the background geometry only through a certain supersymmetric Killing vector field $K$.\footnote{This 
was also argued independently in \cite{Closset:2013vra}.} There are rich classes of $\mathcal{N}=2$ superconformal gauge theories 
which have a large $N$ gravity dual in M-theory. For these theories one can compute the large $N$ 
limit of the partition function using the matrix model saddle point technique of \cite{Herzog:2010hf}. When $M_3$ is diffeomorphic to $S^3$ with the standard action of $U(1)\times U(1)$ on $S^3\subset \R^2\oplus \R^2$, 
and writing $K=b_1\partial_{\varphi_1}+b_2\partial_{\varphi_2}$ in terms of the generators $\partial_{\varphi_i}$ of $U(1)\times U(1)$, one finds 
\cite{Alday:2013lba} the large $N$  free energy $\mathcal{F}=-\log Z$  satisfies
 \bea\label{freeen}
 \lim_{N\to \infty}\,  \frac{{\cal F}}{{\cal F}_{\mathrm{round}} } & = & \frac{(|b_1|+|b_2|)^2}{4|b_1b_2|}  ~,
 \eea
where ${\mathcal F}_{\mathrm{round}}$ is the large $N$  limit of the free energy on the \emph{round} three-sphere, which scales as $N^{3/2}$ \cite{Drukker:2010nc}. 

In \cite{Farquet:2014kma} the field theory result (\ref{freeen}) was reproduced in a dual computation in four-dimensional 
gauged supergravity. Here $M_3\cong S^3$ arises as the conformal boundary of a self-dual Einstein four-manifold $M_4$, where the 
supersymmetric Killing vector $K$ also extends over $M_4$. The asymptotically locally Euclidean AdS metric on $M_4$ is 
conformally K\"ahler, and supersymmetry requires one to turn on a 
 graviphoton field $A$ proportional to the Ricci one-form of this K\"ahler metric. 
A remarkable feature of the computation of the holographic free energy in \cite{Farquet:2014kma} is that 
one does not need to know the form of the Einstein metric on $M_4$ explicitly -- rather (\ref{freeen}) is proven for 
an \emph{arbitrary} such metric. 

In \cite{Farquet:2013cwa} the vacuum expectation values of BPS Wilson loops on the  round sphere were computed for a variety 
of gauge theories,
and  matched to regularized M2-brane actions in AdS$_4\times Y_7$. 
Here the choice of  internal space $Y_7$ determines the gauge theory on $M_3$.
The purpose of this paper is to extend these computations to general supersymmetric backgrounds $M_3=\de M_4$. 
A Wilson loop is BPS if it wraps an orbit of $K$, and 
we will find that the large $N$ Wilson loop VEV satisfies
\bea\label{Wilvevres}
 \lim_{N\to \infty}\, \frac{\log\ \langle \, W\, \rangle}{ \log\ \langle \, W_{\mathrm{round}} \, \rangle} &=& \mfS_{b_1,b_2}\ ,
\eea
where
\bea\label{defSb1b2}
\mfS_{b_1,b_2} & \equiv & \frac{|b_1|+|b_2|}{2}\ell\ .
\eea
Here $\langle \, W_{\mathrm{round}} \, \rangle$ denotes the large $N$ limit of the Wilson loop on the round sphere, whose logarithm scales as $N^{1/2}$, and $2\pi\ell$ denotes the length of the orbit of $K$. Such orbits always close over the poles of $S^3$, {\it i.e.} at the origins of each copy of $\mathbb{R}^2$ in $S^3\subset\mathbb{R}^2\oplus \mathbb{R}^2$, where the lengths are then $\ell=1/|b_1|$ and $\ell=1/|b_2|$, respectively. For these Wilson loops \eqref{Wilvevres} becomes a function of $b_1/b_2$, exactly as in \eqref{freeen}. The supergravity dual configurations are given by M2-branes wrapping a supersymmetric copy of the M-theory circle 
in $Y_7$ \cite{Farquet:2013cwa} and a complex curve $\Sigma_2\subset M_4$, with boundary $\partial\Sigma_2\subset M_3$ being the Wilson line. Identifying the logarithm of the VEV with minus the holographically renormalized
M2-brane action, we also prove that (\ref{Wilvevres}) holds in general, thus verifying the matching 
of this observable in AdS/CFT in a very broad (infinite-dimensional) class of backgrounds.

The outline of the rest of this paper is as follows. In section \ref{WilQFTsec} we review the geometry of $M_3$, the definition of the BPS Wilson loop and how it may be computed using localization techniques in the large $N$ limit to find \eqref{Wilvevres}. Section \ref{BPSM2} analyses supersymmetric M2-branes in $M_4\times Y_7$ backgrounds in M-theory and we also derive the formula \eqref{Wilvevres} in supergravity. Since our arguments are for general backgrounds they are  somewhat implicit; in section 
\ref{examples} we therefore look at some explicit toric self-dual Einstein spaces, to exemplify our general formulae. 
We conclude in section \ref{discuss} with a brief discussion.

\section{Wilson loops in $\mathcal N=2$ gauge theories on $M_3$}\label{WilQFTsec}

The field theories of interest  have UV descriptions as $\mathcal{N}=2$ Chern-Simons gauge theories coupled to matter on $M_3$, where $M_3$ is a supersymmetric three-manifold. We begin this section by reviewing the geometry of $M_3$, and then define the BPS Wilson loops of interest.  These have been studied 
on particular squashed sphere backgrounds in \cite{Kapustin:2013hpk}, \cite{Tanaka:2012nr} (see also 
\cite{Fujitsuka:2013fga}, \cite{Lewkowycz:2013laa}), and the extension to the general backgrounds 
of \cite{Alday:2013lba, Closset:2012ru} is straightforward. After explaining how the Wilson loop 
VEVs localize in the matrix model, we then take the large $N$ limit to derive (\ref{Wilvevres}). 
 
\subsection{Three-dimensional background geometry}\label{3dgeom}

The manifold $M_3$  belongs to a general class of ``real'' supersymmetric backgrounds, with two supercharges related to one another by charge conjugation \cite{Closset:2012ru}. If $\chi$ denotes the Killing spinor on $M_3$ then there is an associated Killing vector field
\bea\label{Kbil}
K & \equiv & \chi^\dagger \gamma^\mu \chi \partial_\mu \ = \ \partial_\psi~.
\eea
This Killing vector is nowhere zero and therefore defines a foliation of the three-manifold. This foliation is transversely holomorphic with local complex coordinate $z$. In terms of these coordinates the background metric may be written as\footnote{More generally there is a conformal 
factor for this metric \cite{Closset:2012ru}. However, as in \cite{Farquet:2014kma} we are interested in conformal field theories 
with gravity duals, and we may hence set this conformal factor to 1.}
\bea\label{metric}
\dd s^2_{M_3} &=&( \dd \psi + \phi_{(0)} )^2 + 4 \ex^{w_{(0)}} \dd z \dd \bar z \, ,
\eea
where $\phi_{(0)}= \phi_{(0)} (z,\bar{z}) \dd z + \overline{\phi_{(0)} (z,\bar{z} )} \dd \bar z$ is a local one-form and $w_{(0)}(z,\bar z)$ is a function. We introduce an orthonormal frame for the three-metric $\dd s^2_{M_3}$:
\bea\label{3frame}
e^1_{(3)} \ = \ \dd \psi + \phi_{(0)}~,\hspace {12 mm}  e_{(3)}^2+\ii e_{(3)}^3 \ = \ 2 \ex^{w_{(0)}/2} \dd z~,
\eea
and will use indices $i,j,k=1,2,3$ for this frame. 

It is important to stress here that \emph{arbitrary} choices for $\phi_{(0)}$ and 
$w_{(0)}$ (subject to $M_3$ being smooth) lead to supersymmetric backgrounds.
The corresponding Killing spinor equation for $\chi$ may be found in \cite{Alday:2013lba, Closset:2012ru}. Choosing the three-dimensional gamma matrices, in the frame (\ref{3frame}), to be simply the 
Pauli matrices, one finds that the 
Killing spinor solution is
\bea\label{chi}
\chi &= &\ex^{\ii\alpha(\psi,z,\bz)}  \left( \begin{array}{c} \chi_0 \\ \chi_0 \end{array} \right)~,
\eea
where $\chi_0$ is a constant and $\alpha(\psi,z,\bz)$ is a phase. The latter will play an important 
role later.

In much of what follows, and as in \cite{Alday:2013lba}, we will assume that $M_3\cong S^3$ with a toric  structure, so that we have a $U(1)\times U(1)$ symmetry. If we realize $M_3\cong S^3\subset \R^2\oplus\R^2$ then we may write
\bea\label{Kil}
K &=& b_1\partial_{\varphi_1}+b_2\partial_{\varphi_2}~,
\eea
where $\varphi_1$, $\varphi_2$ are standard $2\pi$-period coordinates on $U(1)\times U(1)$.

\subsection{The Wilson loop}
 In $\mathcal{N}=2$ supersymmetric gauge theories the gauge field $\mathcal{A}_i$ is part of a vector multiplet that also contains two real scalars $\sigma$ and $D$ and a two-component spinor $\lambda$, all of which are in the adjoint representation of the gauge group $G$. 
The BPS Wilson loop in a representation $\mathcal{R}$ of  $G$ is given by
\bea\label{defWil}
W & =& \frac{1}{\dim \mathcal{R}}\Tr_\mathcal{R}\left[\mathcal{P}\exp \left(\oint_\gamma \dd s(\ii \mathcal{A}_i\dot x^i+\sigma|\dot x|)\right)\right]~,
\eea
where $x^i(s)$ parametrizes the worldline $\gamma\subset M_3$ of the Wilson loop and the path ordering operator has been denoted by $\mathcal{P}$. 
For a Chern-Simons theory the gauge multiplet has a kinetic term described by the supersymmetric Chern-Simons action
\be\label{CS}
S_{\mathrm{CS}} \ = \ \frac{\ii k}{4\pi}\int\Tr\left[\mathcal{A}\wedge\dd \mathcal{A}+\frac{2}{3}\mathcal{A}\wedge \mathcal{A}\wedge \mathcal{A}+(2D\sigma-\lambda^\dag\lambda)\vol_3\right]~,
\ee
where $k$ denotes the Chern-Simons coupling and $\vol_3$ is the Riemannian volume form on $M_3$. 

The full set of supersymmetry transformations for a vector multiplet and matter multiplet may be found in \cite{Alday:2013lba}.
For our purposes we need note only that localization of the path integral, discussed in the next section, requires 
one to choose a Killing spinor, namely $\chi$ in (\ref{chi}). We then need the following two 
supersymmetry transformations
\bea\label{SUSY}
\nonumber\delta \mathcal{A}_i&=& -\frac{\ii}{2}\lambda^\dag\tau_i\chi~,\qquad 
\delta\sigma\ = \ -\frac{1}{2}\lambda^\dag\chi~,\eea
where $\tau_i$ are the Pauli matrices. If one varies the Wilson loop \eqref{defWil} under the latter supersymmetry transformation one obtains
\bea
\delta W  & \propto & \frac{1}{2}\lambda^\dag(\tau_i\dot x^i-|\dot x|)\chi~.
\eea
The Wilson loop is then invariant under supersymmetry provided
\bea\label{BPSloop}
(\tau_i\dot x^i-|\dot x|)\chi & = & 0~.
\eea
Choosing $s$ to parametrize arclength, so that $|\dot x|=1$ along the loop, it is straightforward to show that (\ref{BPSloop}) 
is satisfied if and only if the Wilson loop  lies along the $e^1_{(3)}$ direction. From (\ref{3frame}) we see that $e^1_{(3)}$ is the one-form dual to the supersymmetric Killing vector $K=\partial_\psi$.
Thus the Wilson loop \eqref{defWil} is indeed a BPS operator provided one takes $\gamma$ to be an orbit of $K$. Notice that the topology of $M_3$ has not been used in this subsection, and thus any Wilson loop wrapped along an orbit of $K$ is BPS, regardless of the topology of $M_3$.

\subsection{Localization in the matrix model}

The VEV of the BPS Wilson loop (\ref{defWil}) is, by definition, obtained by inserting $W$ into the path integral 
for the theory on $M_3$. The computation of this is  greatly simplified by the fact that this path integral 
\emph{localizes} onto supersymmetric configurations of fields. This is by now a fairly standard computation, 
and we shall simply summarize the main steps, referring the reader to \cite{Alday:2013lba, Herzog:2010hf, Kapustin:2009kz, Jafferis:2010un, Martelli:2011qj}
for further details. In particular the localization of the Wilson loop was explained in detail in \cite{Farquet:2013cwa} for the round $S^3$ case. This section  generalizes that discussion to a generic supersymmetric manifold $M_3\cong S^3$.

The central idea is that the path integral, with $W$ inserted, is invariant under the supersymmetry variation 
$\delta$ corresponding to the Killing  spinor $\chi$. We have written two 
of the supersymmetry variations in (\ref{SUSY}), and the variations of other fields (including fields in the chiral matter multiplets) 
may be found on the curved background $M_3$ in \cite{Alday:2013lba}.
 Crucially, $\delta^2=0$ is nilpotent. There is then a form of fixed point theorem
that implies that the only net contributions to this path integral come from field configurations 
that are invariant under $\delta$ \cite{Witten:1991zz}.

For the $\mathcal{N}=2$ supersymmetric Chern-Simons-matter theories of interest, one finds that 
the $\delta$-invariant configurations on $M_3\cong S^3$ are particularly simple: 
\bea\label{localized}
\mathcal{A}_i &  =& 0~, \quad\sigma \ = \ \text{constant}~, \quad  D \ =\  - \sigma h~,
\eea
where the function $h=\frac{1}{2}*(e^1_{(3)}\wedge \dd e^1_{(3)})$, and 
with all fields in the matter multiplet set identically to zero \cite{Alday:2013lba}.  Here we may diagonalize 
$\sigma$ by a gauge transformation. The exact localized partition function then takes the saddle point form \cite{Alday:2013lba}
\bea
\label{Zloc}
Z \, = \, \int \dd\sigma \, \ex^{-\frac{\ii\pi k}{|b_1b_2|}\mathrm{Tr}\, \sigma^2}\prod_{\alpha\in \Delta_+} 4
\sinh \frac{\pi\sigma\alpha}{|b_1|}\sinh\frac{\pi\sigma\alpha}{|b_2|}\prod_{\rho}s_\beta\left[\frac{\ii Q}{2}(1-r)-\frac{\rho(\sigma)}{\sqrt{|b_1b_2|}}\right]~.
\eea
Here the integral is over the Cartan of the gauge group, $k$ denotes the Chern-Simons level, the first product is over positive roots $\alpha\in\Delta_+$ of the gauge group, and the second product is over weights $\rho$ in the weight space decomposition for a chiral matter field in an arbitrary representation $\mathcal{R}_{\mathrm{matter}}$ of the gauge group. We have also defined 
\bea
\beta &\equiv & \sqrt{\left|\frac{b_1}{b_2}\right|}~, \qquad Q \ \equiv \ \beta + \frac{1}{\beta}~,
\eea
the R-charge of the matter field is denoted $r$, and $s_\beta(z)$ denotes the double sine function. 

In this set-up, the VEV of the BPS Wilson loop (\ref{defWil}) reduces to
\bea\label{Wloc}
\langle \, W\,  \rangle \ =\ \frac{1}{Z\dim\mathcal{R}}\int &\dd\sigma& \ex^{-\frac{\ii\pi k}{|b_1b_2|}\mathrm{Tr}\, \sigma^2}\prod_{\alpha\in \Delta_+} 4
\sinh \frac{\pi\sigma\alpha}{|b_1|}\sinh\frac{\pi\sigma\alpha}{|b_2|}\nn\\
&&\times \prod_{\rho}s_\beta\left[\frac{\ii Q}{2}(1-r)-\frac{\rho(\sigma)}{\sqrt{|b_1b_2|}}\right]  \mathrm{Tr}_{\mathcal{R}} \left(\ex^{2\pi \ell \sigma}\right) ~.
\eea
Notice the integrand is the same as that for the partition function (\ref{Zloc}), with an additional insertion 
of $\mathrm{Tr}_\mathcal{R}(\ex^{2\pi\ell\sigma})$ arising from the Wilson loop operator. Note also that we have normalized the VEV relative to the partition function $Z$, so that $\langle \, 1 \, \rangle =1$, as is usual in quantum field theory. We have also defined
\bea\label{length}
\oint_\gamma \dd s \ = \ 2\pi\ell\,
\eea
so that $\ell$ parametrizes the length of the Wilson line. More precisely, the 
 integral  (\ref{length}) is well-defined only for a closed orbit of 
the Killing vector $K=\de_\psi=b_1\de_{\varphi_1}+b_2\de_{\varphi_2}$. 
A generic orbit is closed only when $b_1/b_2\in\mathbb{Q}$ is rational, so that $K$ generates 
a circle subgroup of $U(1)\times U(1)$. Writing $b_1/b_2=m/n$ with $m, n \in \mathbb{Z}$ 
relatively prime integers, these define torus knots via $\gamma\subset T^2\subset S^3$, 
where the homology class $[\gamma]=(m,n)\in H_1(T^2,\Z)\cong \Z\oplus \Z$. These have been studied in the 
present context in \cite{Tanaka:2012nr}. 
If on the other hand $b_1/b_2$ is irrational, 
then the \emph{only} closed orbits are at the two ``poles'' of $M_3\cong S^3$, where $\de_{\varphi_1}=0$ and $\de_{\varphi_2}=0$, respectively. 
Over these poles  
$\oint_\gamma \dd s=2\pi/|b_{2}|$, $2\pi/|b_{1}|$, respectively. Wherever the loop is located, we denote its length $\oint_{\gamma}\dd s$ by $2\pi \ell$ as above.

For a $U(N)$ gauge group we may write
$\sigma=\mathrm{diag}(\frac{\lambda_1}{2\pi},\ldots
\frac{\lambda_N}{2\pi})$,  thus parametrizing $2\pi\sigma$ by its eigenvalues $\lambda_i$. 
Localization has then reduced the partition function $Z$ and the Wilson loop VEV 
to finite-dimensional integrals  (\ref{Zloc}), (\ref{Wloc})  over these eigenvalues,  
but in practice the formulae are difficult to evaluate explicitly. For comparison to the dual supergravity 
results we must take the $N\rightarrow\infty$ limit, where the number of eigenvalues, 
and hence integrals, tends to infinity. One can then attempt to compute this 
limit using a saddle point approximation of the integral. In \cite{Herzog:2010hf} a simple 
\emph{ansatz} for the large $N$ limit of the saddle point eigenvalue distribution 
was introduced. One seeks saddle points with eigenvalues of the form
\bea\label{ansatzl}
\lambda_i & = & x_i N^{1/2}+\ii y_i~,
\eea
with $x_i$ and $y_i$ real and assumed to be $\mathcal{O}(1)$ in a large $N$ expansion. In the large $N$ 
limit the real part is assumed to become dense. Ordering the eigenvalues so that the $x_i$ are strictly increasing, 
the real part becomes a continuous variable $x$, with density $\rho(x)$, while $y_i$ becomes a continuous function of $x$, $y(x)$.

Writing $Z=\ex^{-\mathcal F}$ one then obtains a functional $\mathcal F[\rho(x),y(x)]$, with 
$x$ supported on some interval $[\xmin,\xmax]$, and to apply the saddle point 
method one then extremizes $\mathcal F$ with respect to $\rho(x)$, $y(x)$, subject
to the constraint that $\rho(x)$ is a density
\bea\label{density}
\int_{\xmin}^{\xmax}\rho(x)\dd x &=& 1~.
\eea
One then finally also extremizes over the choice of interval, by varying 
with respect to $\xmin$, $\xmax$, to obtain the saddle point 
eigenvalue distribution $\rho(x)$, $y(x)$.

As it turns out, if one caries out the large $N$ limit with the ansatz \eqref{ansatzl}, one finds a very simple relation between the round sphere results  $\mathcal F_\mathrm{round}$ and $\log\ \langle \, W_{\mathrm{round}}\, \rangle$ and their squashed counterparts (with arbitrary $b_1$ and $b_2$) $\mathcal F$ and $\log\ \langle \, W \, \rangle$. To obtain this result for $\mathcal F$, one may first relabel $\sigma$ as $|b_2| \sigma$ in \eqref{Zloc}. 
The partition function then takes the same form as that in  \cite{Martelli:2011fu}, where the large $N$ limit was computed in detail.
In particular in the latter reference it was shown that in the large $N$ limit $\mathcal F[\rho(x),y(x)]$ is simply a rescaling of the round sphere 
result by a factor $(\beta Q)^3/2^3\beta^2$, \emph{provided} one also rescales the Chern-Simons coupling $k$ as 
$k\rightarrow (2/\beta Q)^2\cdot k$. This then leads to the large $N$ result (\ref{freeen}). 

The same logic may be applied to the calculation of the Wilson loop.
For the class of $\mathcal{N}=2$ supersymmetric Chern-Simons theories coupled to  matter on the round 
three-sphere studied in \cite{Farquet:2013cwa}, $x_\text{max}$ is always proportional to $1/\sqrt{k}$. 
According to the above prescription, the result for $x_\text{max}$ on a general background $M_3$ is given by rescaling the round sphere result by 
$|b_2|\cdot (\beta Q/2) = (|b_1|+|b_2|)/2$. Here the factor of $|b_2|$ comes from the relabelling $\sigma\rightarrow |b_2|\sigma$, while 
the factor of $\beta Q/2$ comes from the rescaling of the Chern-Simons coupling. Thus
\bea
\xmax & =& \frac{|b_1|+|b_2|}{2}\ x_\text{max}^\text{round}\ ,
\eea
where $x_\text{max}^{\text{round}}$ determines the supremum of the support of $\rho(x)$ for the field theory 
on the round three-sphere. For the theories studied in 
\cite{Farquet:2013cwa}, the eigenvalue density is always a  continuous piecewise linear function supported on $[\xmin,\xmax]$. Using this fact, the large $N$ limit of the Wilson loop \eqref{Wloc} in the fundamental representation may be easily computed with a saddle point approximation, and is 
\bea\label{Wbint}
\log\ \langle \, W\, \rangle_\text{QFT}\ = \ \ell\cdot x_\text{max}\ N^{1/2}+o(N^{1/2})\ .
\eea
Here recall that the length $\oint_\gamma \dd s$ is in general $2\pi\ell$. 
The round three-sphere Wilson loop in particular is obtained by setting $b_1=b_2=1$ and $\ell=1$ and is, as shown in \cite{Farquet:2013cwa}, 
\bea
\log\ \langle \, W_{\mathrm{round}} \, \rangle_\text{QFT}\ = \ x^{\text{round}}_\text{max}\ N^{1/2}+o(N^{1/2})\ .
\eea
We thus obtain
\bea\label{loopQFT}
\lim_{N\rightarrow\infty}\, \frac{\log\  \langle  \, W\, \rangle_\text{QFT}}{\log\ \, \langle W_{\mathrm{round}}\, \rangle_\text{QFT}}\ =\ \frac{|b_1|+|b_2|}{2}\ell \ .
\eea
This is the field theory result for the VEV of a supersymmetric Wilson loop on a general supersymmetric manifold $M_3\cong S^3$. In the next section we will look at the M2-brane dual to this Wilson loop, and show quite generally that the holographic dual computaton of the VEV agrees with 
(\ref{loopQFT}).

\section{Dual M2-branes}\label{BPSM2}

In this section we analyse the supersymmetric M2-brane probes that are relevant for computing the holographic dual of the Wilson loop VEV \eqref{loopQFT}. The dual solution is constructed in four-dimensional gauged supergravity \cite{Farquet:2014kma}, 
and we begin by summarizing the geometry of these solutions. We then  look at the eleven-dimensional uplift, and finally we compute the regularized action of the M2-brane. 

\subsection{Four-dimensional supergravity dual}\label{4dgeom}

In \cite{Farquet:2014kma} it was shown that supersymmetric three-manifolds $M_3$ of precisely the form described in section \ref{3dgeom} arise 
as the conformal boundaries of Euclidean self-dual solutions to four-dimensional gauged supergravity. For $M_3\cong S^3$ the 
four-dimensional supergravity solution is defined on a four-ball $M_4\cong B^4$, and is asymptotically locally Euclidean AdS with conformal 
boundary $M_3$. The Killing vector $K$ defined by (\ref{Kbil}) extends as a Killing vector bilinear over $M_4$, and the four-metric 
is then Einstein, has anti-self-dual Weyl tensor, and is conformal to a K\"ahler metric. Supersymmetry also requires one to turn on 
a specific graviphoton field $A$. After summarizing these solutions, and deriving some relevant formulae, we then use them to study the BPS M2-branes dual to the Wilson loops 
of the previous section.

The four-dimensional metric on the manifold $M_4$ takes the form
\bea\label{4SDE}
\dd s_{M_4}^2 &= &\frac1{y^2} \left[V^{-1} (\dd \psi+\phi )^2 + V(\dd y^2 + 4\ex^w \dd z \dd \bar z )\right]\ ,
\eea
where 
\bea
V &=& 1 - \frac{1}{2}y\partial_y w~,\label{V}\nn \\
\dd\phi &=& \ii \partial_zV\dd y \wedge \dd z - \ii \partial_{\bar{z}}V\dd y \wedge \dd\bar{z} + 2\ii \partial_y(V\ex^w)\dd z\wedge \dd\bar{z}~, \label{4dphi}
\eea
and $w=w(y,z,\bar{z})$ satisfies  the Toda equation
\bea\label{Todaeq}
\partial_z\partial_{\bar{z}} w + \partial^2_y\ex^w &=& 0~.
\eea
The metric (\ref{4SDE}) is equipped with the Killing vector $K=\de_\psi$, which extends the vector (\ref{Kbil}) on the conformal boundary, which 
is at $y=0$. The coordinate $y$ may be regarded as a radial coordinate, $y\in (0,y_0]$, with the conformal boundary at $y=0$ and 
the origin of $M_4\cong B^4$ being at $y=y_0>0$. The local complex coordinate $z$ similarly extends that on the conformal boundary $M_3$. 
The metric (\ref{4SDE}) is then entirely determined by the solution $w=w(y,z,\bar{z})$ to the Toda equation (\ref{Todaeq}). 

Supersymmetry requires that the graviphoton gauge field $A$ takes the \emph{local} form
\bea\label{Agauge}
 A &=& -\frac{1}{4}V^{-1}\partial_y w(\dd\psi+\phi) +\frac{\ii}{4}\partial_zw\dd z - \frac{\ii}{4}\partial_{\bar{z}}w\dd \bar{z}~,
\eea
where the field strength $F$ is defined by $F=\dd A$. Indeed, the metric (\ref{4SDE}) is conformal 
to the K\"ahler metric $\dd s^2_{\mathrm{Kahler}}=y^2\dd s^2_{M_4}$, which is asymptotic 
to a cylinder $\R\times M_3$ near to the conformal boundary $y=0$. The gauge field (\ref{Agauge}) is 
then $\tfrac{1}{2}$ of the Ricci one-form for this K\"ahler metric.
These solutions were referred to as \emph{self-dual} solutions in \cite{Farquet:2014kma}, 
since the Weyl tensor is anti-self-dual\footnote{With respect to the
canonical orientation of the conformal K\"ahler metric.} and $F$ is anti-self-dual, {\it i.e. } $*_4 F=-F$. 
Moreover, the metric (\ref{4SDE}) is Einstein with negative cosmological constant.
We shall use the following orthonormal frame for \eqref{4SDE}
\bea
e^0 \ = \ \frac{1}y V^{1/2} \dd y\ ,\quad e^1 \ = \ \frac{1}{y} V^{-1/2} (\dd \psi + \phi )\ , \quad e^2+\ii e^3 \ = \ \frac{2}y (Ve^{w})^{1/2} \dd z\ . \label{fra2}
\eea

As already mentioned, the  solutions of interest are asymptotically locally Euclidean AdS (asymptotically hyperbolic), with the conformal boundary at $y=0$. 
In particular imposing boundary conditions such that $w(y,z,\bz)$ is analytic around $y=0$, {\it i.e.}
\bea
w(y,z,\bar{z}) &=& w_{(0)}(z,\bar{z})+y w_{(1)}(z,\bar{z})+\frac{1}{2}y^2w_{(2)}(z,\bar{z})+\mathcal{O}(y^3)~,
\eea
then setting $r=1/y$ the metric \eqref{4SDE} expands to leading order as
\bea\label{asy}
\dd s^2_{M_4} & \simeq & \frac{\dd r^2}{r^2} + r^2\left[(\dd \psi+\phi_{(0)})^2 + 4\ex^{w_{(0)}}\dd z\dd\bar{z}\right]~,
\eea
when $r\to\infty$. Here we have also expanded the one-form $\phi$ tangent to $M_3$
\bea
\phi (y,z,\bar z)\mid_{TM_3}  & = & \phi_{(0)}(z,\bar z) + y \phi_{(1)}(z,\bar z) + \mathcal O(y^2)\label{phiexp}.
\eea
In fact by expanding (\ref{4dphi}) one can show that $\phi_{(1)}=0$. 
Equation (\ref{asy}) shows explicitly that the metric is asymptotically locally Euclidean AdS around $y=0$, and moreover 
there is a natural choice of conformal class for the metric on the boundary $M_3$ given  precisely by (\ref{metric}).

 The four-dimensional geometry that we have just described, together with the gauge field $A$, form a supersymmetric solution to Euclidean gauged supergravity. 
There is correspondingly a Dirac spinor $\epsilon$ satisfying the Killing spinor equation of this theory. In the orthonormal frame
(\ref{fra2}) and using the gamma matrices
\bea\label{Gammas}
\Gamma_i & =& \left(\begin{array}{c c} 0 & \tau_{i} \\  \tau_{i} & 0 \end{array} \right)~, \hspace{10 mm} \Gamma_0 \ =\ \left(\begin{array}{c c} 0 & \ii \mathbb I_2 \\ -\ii \mathbb I_2 & 0 \end{array} \right)~,
\eea
with $\tau_i$ the Pauli matrices, the Killing spinor $\epsilon$ is given by
\bea\label{diracspin4}
\epsilon & =&  \frac{1}{\sqrt{2y}}\left(1+V^{-1/2}\Gamma_0\right)\zeta~.
\eea
with
\bea\label{zeta0}
\zeta & =& \begin{pmatrix}\chi\\ 0\end{pmatrix}\quad\text{where}\quad \chi \ = \ \begin{pmatrix}\chi_0\\ \chi_0\end{pmatrix}~.
\eea
In particular 
the bulk spinor (\ref{diracspin4}) precisely matches onto the boundary two-component spinor $\chi$ given by (\ref{chi}). 
The phase in (\ref{chi}) may be shifted locally by making gauge transformations of $A$, since the Killing spinor is 
charged under the latter. However, for these solutions one may write $A$ as a \emph{global} one-form on $M_4$. This requires making an appropriate gauge transformation on the local expression 
(\ref{Agauge}), as we shall see in the next subsection.

So far we have not imposed the $U(1)\times U(1)$ symmetry we imposed on the boundary $M_3$ at the end of section \ref{3dgeom}. 
Doing so will simplify the subsequent discussion. Thus as in 
 \cite{Farquet:2014kma} we assume that the four-manifold $M_4$ is $M_4\cong B^4\cong \mathbb{R}^2\oplus \mathbb{R}^2$ 
and that the torus $U(1)\times U(1)$ acts in the standard way on $\R^2\oplus \R^2$. 
The Killing vector $K=\partial_\psi$ is then 
parametrized as
\bea\label{Kagain}
K &=& b_1\partial_{\varphi_1}+b_2\partial_{\varphi_2}~,
\eea
again precisely as in (\ref{Kil}) on the conformal boundary. 
It will be important to fix carefully the orientations here.
Since the metrics are defined on a ball, diffeomorphic to $\R^4\cong \R^2\oplus \R^2$ 
with $U(1)\times U(1)$ acting in the obvious way, we choose $\partial_{\varphi_i}$ 
so that the orientations on $\R^2$ induce the given orientation on $\R^4$ 
(with respect to which the metric has anti-self-dual Weyl tensor). This fixes 
the relative signs of $b_1$ and $b_2$. Given that 
$K$ has no fixed points near the conformal boundary, we must also have $b_1$ and 
$b_2$ non-zero. Thus $b_1/b_2\in \R\setminus\{0\}$, and its sign will be important 
in what follows. 

In order to construct such backgrounds one can start with a toric ($U(1)\times U(1)$-invariant) 
self-dual Einstein metric on a ball, which is asymptotically locally Euclidean AdS. 
There are many examples of such metrics -- we discuss the two simplest in section \ref{examples}, 
but as explained in \cite{Farquet:2014kma} the moduli space is in fact infinite-dimensional (each metric inducing a different conformal structure on the boundary $M_3\cong S^3$). One can then choose 
a Killing vector (\ref{Kagain}), and then writing $K=\partial_\psi$ the metric will necessarily take the 
form (\ref{4SDE}). Thus in particular the choice of $K$ determines the conformal 
K\"ahler metric, which in turn determines the instanton gauge field $A$ and Killing spinor $\epsilon$. 
However, not all choices of $K$ in (\ref{Kagain}) give non-singular gauge fields. 
While the metric (\ref{4SDE}) is smooth by assumption, the instanton $F=\dd A$ and Killing spinor $\epsilon$ are singular 
where the conformal K\"ahler metric is singular.  Regularity is in fact equivalent to having either $b_1/b_2>0$ or $b_1/b_2=-1$.
Moreover, the origin $y=y_0$ of $M_4\cong \R^2\oplus \R^2$ is then at 
\bea\label{yNUT}
y_0 &=& \frac{1}{|b_1+b_2|}~,
\eea
which notice is $y_0=\infty$ when $b_1/b_2=-1$.

\subsection{Global gauge for $A$}

As remarked after equation (\ref{zeta0}), we will want to choose a gauge for $A$ in which it is a global, smooth one-form on $M_4$. 
The reason for this is that we will evaluate the Wess-Zumino term in the M2-brane action in section \ref{M2sec} by using Stokes' theorem for $F=\dd A$, which 
requires us to write $A$ as a  global one-form.  This was also discussed to some extent in \cite{Farquet:2014kma}, but 
for the computation of the Wilson loop we need a little more information. 

The key point is to recall that $A$ is proportional to the Ricci one-form for the conformal K\"ahler metric 
$\dd s^2_{\mathrm{Kahler}}=y^2\dd  s^2_{M_4}$. When $b_1/b_2>0$ the associated complex 
structure identifies $M_4\cong \R^2\oplus\R^2\cong\C^2$. The orientation in which the Weyl tensor is anti-self-dual is  
the same as the canonical orientation on $\C^2$. One can then introduce standard complex coordinates 
$z_i=\rho_i\ex^{\ii\psi_i}$, $i=1,2$, on $\C^2$. The spinor $\zeta$ in (\ref{zeta0}), which is used to construct 
the Killing spinor (\ref{diracspin4}), is the canonical spinor that exists on any K\"ahler manifold \cite{Farquet:2014kma}. 
As such we have
\bea
\mathcal{L}_{\partial_{\psi_i}} \epsilon &=& \frac{\ii}{2}\epsilon~, \qquad i \ = \ 1,2~.
\eea
Denoting the complex structure tensor by $J$ we also have that
$J(V^{-1}\partial_y)=\partial_\psi=K$. Since $y$ is \emph{decreasing} as we move away 
from the origin of $\C^2$, where recall that the origin is at $y=y_0>0$,
this means that for $b_1>0$ and $b_2>0$ we 
must then identify $\varphi_i=-\psi_i$, where $\varphi_i$ are the coordinates 
on $U(1)\times U(1)$ in (\ref{Kagain}). This is because for $r$ any radial coordinate 
on $\C^2$ we have $J(r\partial_r)=a_1\partial_{\psi_1}+a_2\partial_{\psi_2}$ 
where necessarily $a_1, a_2>0$ (that is, the Reeb cone is the positive quadrant in $\R^2$ -- see, for example, \cite{Sparks:2010sn}). On 
the other hand for $b_1<0$ and $b_2<0$ we instead have $\varphi_i=+\psi_i$, $i=1,2$.

The other non-singular case is $b_1/b_2=-1$. This is qualitatively different from the case $b_1/b_2>0$ in the 
last paragraph, as here $y_0=\infty$ (\ref{yNUT}). Moreover, the origin $y=y_0$ of $M_4\cong \R^2\oplus\R^2$ 
is now identified with the point at infinity in $\C^2$, rather than the origin. One can see this from the conformal K\"ahler metric 
$\dd s^2_{\mathrm{Kahler}}=y^2\dd  s^2_{M_4}$, which is asymptotically Euclidean around $y=y_0$. 
Thus now $V^{-1}\partial_y$ has the correct orientation for a radial vector on $\C^2$, and we 
deduce that for $b_1<0$ and $b_2>0$ we have $\varphi_1=-\psi_1$, $\varphi_2=+\psi_2$, while 
for $b_1>0$ and $b_2<0$ we instead have $\varphi_1=+\psi_1$, $\varphi_2=-\psi_2$.

Putting all of the above together, we may compute the charge of the Killing spinor $\epsilon$ under the supersymmetric Killing vector 
$K=\partial_\psi$:
\bea\label{charge}
\mathcal{L}_K\epsilon &=& \ii \gamma \epsilon~,
\eea
where\footnote{We also denoted the Wilson line curve by $\gamma:S^1\rightarrow M_3$ in section \ref{WilQFTsec}, but have 
chosen to use the same symbol for the charge of $\epsilon$ under $K$ as this was also used in \cite{Alday:2013lba}, \cite{Farquet:2014kma}.}
\bea\label{gamma}
\gamma &\equiv & -\mathrm{sign}\left(\frac{b_1}{b_2}\right)\cdot\frac{|b_1|+|b_2|}{2}~.
\eea

Since in all cases the K\"ahler structure is defined on $\C^2$, the canonical bundle is trivial and
one may indeed take $A$ to be a global one-form on $M_4$. We denote the restriction of this global $A$
to the conformal boundary $M_3=\de M_4$ at $y=0$ by $A_{(0)}$. Then the formula (\ref{charge}) 
for the charge of $\epsilon$ under $K=\partial_\psi$ means that
\bea\label{Aglobbound}
A_{(0)} & = & \gamma \dd\psi -\frac{1}{4}w_{(1)}(\dd\psi+\phi_{(0)}) +\frac{\ii}{4}\partial_zw_{(0)}\dd z - \frac{\ii}{4}\partial_{\bar{z}}w_{(0)}\dd \bar{z}~.
\eea
This is the restriction of (\ref{Agauge}) to $y=0$, together with a gauge transformation $A\rightarrow A+\gamma\dd\psi$ which 
accounts for the charge (\ref{charge}). One can show independently that (\ref{Aglobbound}) then defines a global 
one-form on $M_3\cong S^3$, which leads to another formula for $\gamma$ that was derived in section 3.3 of \cite{Farquet:2014kma}, 
although we will not need this in the present paper.

\subsection{Uplifting to $D=11$ supergravity}

In order to study the M2-branes dual to Wilson loops, we need to uplift the four-dimensional geometry to an eleven-dimensional supergravity solution. More precisely, we are interested in a class of $\mathcal{N}=2$ supersymmetric $M_4\times Y_7$ backgrounds of M-theory in \emph{Euclidean} signature. In Euclidean signature there are certain factors of $\ii$ that appear relative to the uplifting formula in Lorentzian signature 
of \cite{Gauntlett:2007ma}. Again, this will be important for correctly evaluating the M2-brane action.

 The action of $D=11$ supergravity in Euclidean  signature is 
\bea
S_{11}\ =
 \ \frac{1}{(2\pi)^8 \ell_p^9 }\left( \int \dd^{11} x \sqrt{{ g_{11}}} \left[- \mathcal{R}+\frac{1}{2}\dd C\wedge *_{11}\dd C\right]+  \frac{\ii}{6} \int C\wedge \dd C\wedge \dd C\right)\ .
\eea
Here we have denoted by ${g_{11}}$ the eleven-dimensional metric, with associated Ricci scalar $\mathcal{R}$, $C$ is the three-form potential and $\ell_p$ denotes the eleven-dimensional Planck length. The equations of motion for the metric and $C$-field follow immediately:
\bea\nn\label{11eom}
\mathcal{R}_{AB}-\frac{1}{12}(G_{A C_1C_2C_3}G{_{B}}{^{C_1C_2C_3}}-
\frac{1}{12}g_{AB}G^2)&=&0\ ,\\
\dd *_{11}G+\frac{\ii}{2}G\wedge G&=&0\ ,
\eea
where we have defined $G\equiv \dd C$ and $A,B,C=1,\ldots,11$. It is also useful to define $G_7=\ii(*_{11}G+\frac{\ii}{2}C\wedge G)$ so that the equation of motion for $G$ is simply $\dd G_7=0$. 

An ansatz that leads to a consistent truncation to four-dimensional gauged supergravity in Lorentz signature  was given  in \cite{Gauntlett:2007ma}. 
Here the internal space $Y_7$ is taken to be 
 any Sasaki-Einstein seven-manifold $Y_7$ with contact one-form $\eta$, transverse K\"ahler-Einstein six-metric $\dd s_T^2$ with K\"ahler form $\omega_T=\dd\eta/2$, and with the seven-dimensional metric normalized so that $\mathrm{Ric}=6 g_{Y_7}$. The consistent truncation 
ansatz  in Euclidean signature then becomes
\bea\nn\label{11ansatz}
\dd s^2_{11}&=&R^2\left[\frac{1}{4} \dd s^2_{M_4}+\left(\eta+\frac{1}{2} A\right)^2+\dd s_T^2\right]\ ,\\
G&=&-\ii R^3\left(\frac{3}{8}\vol_4-\frac{1}{4}*_4F\wedge \dd\eta\right)\ .
\eea
As before, $\dd s^2_{M_4}$ is the four-dimensional gauged supergravity metric on $M_4$ with gauge field $A$, field-strength $F=\dd A$ and volume form $\vol_4$. The radius $R$ is 
\bea
R^6 &= & \frac{(2\pi \ell_p)^6 N}{6\Vol(Y_7)}\ ,
\eea
where $N$ is the number of units of flux
\bea\label{DiracN}
N & =& \frac{1}{(2\pi \ell_p)^6}\int_{Y_7}G_7\ .
\eea
Substituting the ansatz \eqref{11ansatz} into the equations of motion \eqref{11eom}, we find the latter are equivalent to the metric $g_{\mu\nu}$ corresponding to $\dd s^2_{M_4}$ and $F$ satisfying
\bea\label{4eom}
R_{\mu\nu} + 3g_{\mu\nu}  &=& 2\left({F_\mu}^{\rho}F_{\nu\rho}-\frac{1}{4}F^2 g_{\mu\nu}\right)~,\nonumber\\
\dd *_4F &=&0~. 
\eea
The ansatz \eqref{11ansatz} 
then solves the eleven-dimensional Euclidean equations of motion if and only if 
the four-dimensional metric $g_{\mu\nu}$ and gauge field $A$ are a solution of four-dimensional Euclidean gauged supergravity. 

\subsection{BPS M2-branes}\label{M2sec}

We are interested in calculating the action of M2-branes that are dual to Wilson loops of the dual gauge theory on $M_3$. These M2-branes wrap $\Sigma_2\times S^1_M$, where the surface $\Sigma_2\subset M_4$ 
has boundary given by the Wilson line $\partial \Sigma_2=S^1\subset M_3=\partial M_4$, and $ S^1_M\subset Y_7$ is a copy of the M-theory circle. In particular we will show that submanifolds 
 $\Sigma_2\subset M_4$ parametrized by the radial direction $y$ in $M_4$ and an orbit of the Killing vector $K$ are
complex with respect to the complex structure $J$ of the conformal K\"ahler metric to $\dd s^2_{M_4}$. The wrapped M2-brane is 
then supersymmetric.\footnote{More precisely the copy of $S^1_M\subset Y_7$ must also be calibrated by the contact one-form $\eta$ on 
$Y_7$. Since this internal space geometry is identical to the 
 AdS$_4\times Y_7$ backgrounds studied in \cite{Farquet:2013cwa}, in this paper we focus on the geometry of $M_4$. } Over the poles $S^1\subset M_3\cong S^3$ the topology of $\Sigma_2$ is a disc, where $y\in(0,y_0]$ serves as a radial coordinate with the origin of the disc at $y=y_0>0$.

The action of the M2-brane is
\bea\label{actionM2}
S_{\text{M2}} & =& \frac{1}{(2\pi)^2\ell_p^3}\left[\Vol(\Sigma_2\times S^1_M)+\ii\int_{\Sigma_2\times S^1_M} C \right]\ .
\eea
A supersymmetric M2-brane satisfies an appropriate $\kappa$-symmetry condition, which may be written as
 \cite{Becker:1995kb}
\bea\label{kappa}
\mathbb{P}\e_{11}\ =\ 0~,\quad \mbox{where}\quad \mathbb{P} \ \equiv \ \frac{1}{2}\left(1-\frac{\ii}{3!}\varepsilon^{ijk}\de_iX^M\de_jX^N\de_kX^P\hat\Gamma_{MNP}\right)~,
\eea
with $i,j,k$ indices on the worldvolume. Here $\epsilon_{11}$ is the eleven-dimensional Killing spinor for the background (\ref{11ansatz}), 
which is constructed as a tensor product of the four-dimensional spinor $\epsilon$ and the Killing spinor on the internal space $Y_7$. 
The $\hat{\Gamma}_M$ are eleven-dimensional gamma matrices, with $X^M$ describing the M2-brane embedding. 
One can analyse (\ref{kappa}) precisely as the authors did in \cite{Farquet:2013cwa}. The upshot is that  $S^1_M\subset Y_7$ 
must be a calibrated circle in $Y_7$ \cite{Farquet:2013cwa}, while taking $\Sigma_2\subset M_4$ to be a surface at constant 
$z$, parametrized by $y$ and $\psi$, one finds (\ref{kappa}) is equivalent to the projection condition
\be\label{projeps}
(1-\ii \Gamma_5\Gamma_{01})\epsilon\ =\ 0\ .
\ee
Here we have used the orthonormal frame (\ref{fra2}), and $\Gamma_5\equiv\Gamma_0 \Gamma_1\Gamma_2\Gamma_3$ with 
$\Gamma_\mu$ defined by (\ref{Gammas}) (in the orthonormal frame). Using the explicit form for $\epsilon$ in (\ref{diracspin4}) 
it is trivial to see that (\ref{projeps}) indeed holds. Moreover, $\Sigma_2$ is calibrated with respect to the K\"ahler form 
for the conformal K\"ahler metric, making it a complex curve.

Let us now calculate the action \eqref{actionM2} for our M2-brane. Using the self-dual four-dimensional supergravity solution of section \ref{4dgeom} and the uplift \eqref{11ansatz} the $C$-field is computed to be
\bea
C & =& -\ii R^3\left(-\frac{1}{8}\Gamma+\frac{1}{4}F\wedge \eta\right)~,
\eea
where
\bea
\Gamma & \equiv & \frac{1}{2y^2}(\dd\psi+\phi)\wedge \dd\phi + \frac{1}{y^3}(\dd\psi+\phi)\wedge 2\ii V\ex^w \dd z\wedge \dd\bar{z}~,
\eea
and $\dd\Gamma=-3\vol_4$. The area of the surface $\Sigma_2$ in $M_4$ is divergent, but can be regularized by subtracting the length of its boundary, {\it i.e.} the length of the $S^1$ in $M^\delta_3$ at $y= \delta\to 0$. Notice this is then a local boundary counterterm. If we denote by $M_4^\delta$ the manifold $M_4$ with boundary $M_3^\delta = \{y=\delta\}$ (with $0<\delta<y_0$), and similarly for $\Sigma_2^\delta$ {\it etc}, the action of the M2-brane is
\be\label{actionM2gen}
S_{\mathrm{M2}} \ = \ \frac{1}{(2\pi)^2\ell_p^3}\int_{S^1_M}\frac{R^3}{4}\ \vol_{S^1_M}\cdot\lim_{\delta\to 0}\left[\int_{\Sigma_2^\delta}\vol_{\Sigma_2}-\int_{\de\Sigma_2^\delta} e^1_\mu\dd x^\mu +\int_{\Sigma_2^\delta}F\right]\ .
\ee
Here we have written $\vol_{S^1_M}$ for the volume form on $S^1_M$ induced from the metric $g_7$, and similarly for $\vol_{\Sigma_2}$ and the metric $g_{M_4}$. Applying Stokes' theorem for the gauge field term $F=\dd A$ we then compute\footnote{The sign in front of the gauge field term  arises because $y$ is decreasing towards the boundary of $M_4$, and hence $\dd y $ points inwards from $M_3$. Thus the natural orientation 
of the boundary we take is opposite to that in Stokes' theorem. }
\bea\nn
S_{\mathrm{M2}}&=& \frac{1}{(2\pi)^2\ell_p^3}\int_{S^1_M}\vol_{S^1_M}\cdot\frac{\pi \ell R^3}{2}\ \lim_{\delta\to 0}\left[\left(\int_{\delta}^{y_0}\frac{\dd y}{y^2}-\frac{1}{\delta\sqrt{V(\delta,z, \bar z)}}\right)-\frac{1}{2\pi \ell}\int_{\de\Sigma_2^\delta}A\right]\\
&=&\frac{1}{(2\pi)^2\ell_p^3}\int_{S^1_M}\vol_{S^1_M}\cdot\frac{\pi \ell R^3}{2}\ \left[-\left(\frac{1}{y_0}+\frac{1}{4}w_{(1)}\right)-\frac{1}{2\pi \ell}\int_{\de\Sigma_2}A\right]\ .
\eea
Recall here that $2\pi\ell$ denotes the length of the orbit of $K$, as in (\ref{length}).
The contribution of the M-theory circle $S^1_M$ is exactly the same as for the AdS$_4\times Y_7$ backgrounds studied in \cite{Farquet:2013cwa}, and is expressed in terms of the contact form $\eta$ on $Y_7$ and the Dirac quantized number $N$ of \eqref{DiracN}. The gauge field integral is easily computed, thanks to \eqref{Aglobbound}
\be
\label{intASig2}
\int_{\de\Sigma_2}A \ = \ \int_{\de\Sigma_2}A_{(0)} \ = \ 2\pi \ell\left(-\frac{1}{4}w_{(1)}+\gamma\right)\ .
\ee

Putting everything together, and using the formula (\ref{yNUT})  for $y_0$, we have
\be\label{logWgravint}
\log{\langle \, W\, \rangle}_{\text{gravity}}\ =\ -S_{\mathrm{M2}}\ =\ \ell\left(|b_1+b_2|+\gamma\right)\cdot\frac{(2\pi)^2\int_{S^1_M}\eta}{\sqrt{2\int_{Y_7}\eta\wedge(\dd\eta)^3}}\ N^{1/2}\ .
\ee
Using the round sphere result of \cite{Farquet:2013cwa}
\be
\log{\langle \, W_{\mathrm{round}}\, \rangle}_\text{gravity}\ =\ \frac{(2\pi)^2\int_{S^1_M}\eta}{\sqrt{2\int_{Y_7}\eta\wedge(\dd\eta)^3}}\ N^{1/2}\ ,
\ee
and the formula (\ref{gamma}) for $\gamma$, in both cases $b_1/b_2>0$ and $b_1/b_2=-1$ we obtain
\be\label{Wilsongrav}
\log{\langle \, W\, \rangle}_{\text{gravity}}\ =\ \frac{|b_1|+|b_2|}{2}\ell\cdot \log{\langle \, W_{\mathrm{round}}\, \rangle}_{\text{gravity}}\ .
\ee
In \cite{Farquet:2013cwa} it was shown in numerous families of examples that 
 the large $N$ limit of the Wilson loop on the round three-sphere and the M2-brane in AdS$_4$ have the same VEV, {\it i.e.} $\log{\langle \, W_{\mathrm{round}}\, \rangle}_{\text{QFT}}\ =\ \log{\langle \, W_{\mathrm{round}}\, \rangle}_{\text{gravity}}$ holds to leading order at large $N$. Assuming this to be the case,
 equations \eqref{loopQFT} and \eqref{Wilsongrav} mean 
that we have shown very generally that in the large $N$ limit
\bea
\log{\langle \, W\, \rangle}_{\text{QFT}}\ =\ \log{\langle \, W\, \rangle}_{\text{gravity}}\, 
\eea
where now the field theory is defined on a general class of background three-manifolds $M_3$, with fillings $M_4$ in four-dimensional 
gauged supergravity.  

We conclude this section with two further comments.
Firstly, it is interesting to note that when the orbit of $K$ is one of the poles of $S^3$, where correspondingly $\ell=1/|b_1|$ or $\ell=1/|b_2|$ respectively, the Wilson loops are then functions only of $|b_1/b_2|$, just as for the free energy \eqref{freeen}. Secondly, in the case that $b_1/b_2=m/n$ is rational and the Wilson line 
wraps a generic orbit $\gamma\subset T^2\subset S^3$ ({\it i.e.} not at either pole), 
then the curve $\Sigma_2\subset M_4\cong \C^2$ wrapped by the dual M2-brane is 
the Brieskorn-Pham curve $\{z_1^n=z_2^m\}\subset \C^2$. This follows since supersymmetry 
pairs the orbit of $K$ with its complexification in $M_4\cong\C^2$, meaning that $\Sigma_2$ 
is swept out as  a generic $\C^*$ orbit of $(z_1,z_2)\rightarrow (\lambda^mz_1,\lambda^nz_2)$, with 
$\lambda\in\C^*$. The curve $\{z_1^n=z_2^m\}$ adds the origin in $\C^2$ at $y=y_0$, 
which is a singular point when $m,n>1$, although notice this does not affect our computation 
of the M2-brane action, which is finite. It is well-known that 
 $(m,n)$ torus knots in $S^3$ may be realized  as links 
of the above Brieskorn-Pham curves, and it is interesting to see that this construction is 
realized as the holographic dual of the knot.

\section{Examples}\label{examples}

Our derivation of the formula (\ref{Wilsongrav}) was necessarily somewhat indirect, as we have shown that it holds for a very general (infinite-dimensional) class of solutions. In particular we didn't need to use the explicit form of the solution to the Toda equation (\ref{Todaeq}). 
In this section we illustrate our general results by discussing two explicit families of solutions, where all quantities in the previous section may be written down in closed form. We will focus on the four-dimensional part of the M2-brane calculation, in particular showing how the
 factor $\ell (|b_1|+|b_2|)/2$ in (\ref{Wilsongrav}) arises explicitly in these cases. 
 In order to do so we will use the results of the previous section that allow us to write 
\be
\log{\langle \, W\, \rangle}_{\text{gravity}}\ =\ \mfS_{b_1,b_2}\cdot \log{\langle \, W_{\mathrm{round}}\, \rangle}_{\text{gravity}}\ ,
\ee
where
\be\label{defSb}
\mfS_{b_1,b_2} \ \equiv  \ \frac{1}{2\pi} \left(-\int_{\Sigma_2}\vol_{\Sigma_2}+\int_{\de\Sigma_2} \vol_{\de\Sigma_2} +\int_{\de\Sigma_2}A\right)~.
\ee
Here we cut off $\Sigma_2$ at $y=\delta$, and (\ref{defSb}) is then understood to be the limit $\delta\rightarrow 0$. 
We compute (\ref{defSb}) directly in the examples, confirming that (\ref{Wilsongrav}) indeed holds in these cases.

\subsubsection*{AdS$_4$}

We begin with the metric on Euclidean AdS$_4$, which can be written 
\bea\label{EAdS}
\dd s^2_{\mathrm{EAdS}_4} &=& \frac{\dd q^2}{q^2+1}+q^2\left(\dd\ppsi^2 + 
\cos^2\ppsi \dd\varphi_1^2+\sin^2\ppsi\dd\varphi_2^2\right)~.
\eea
Here $q$ is a radial variable with  $q\in[0,\infty)$, so that the origin of $M_4\cong \R^4$
is at $q=0$ while the conformal boundary is at $q=\infty$. 
The coordinate $\ppsi\in[0,\frac{\pi}{2}]$, with the endpoints being the two 
axes of $\R^2\oplus\R^2\cong \R^4$. 

Of course the metric (\ref{EAdS}) is conformally flat, which leads to a trivial graviphoton $A=0$. However, we may instead pick a general 
supersymmetric Killing vector  $K=b_1\partial_{\varphi_1}+b_2\partial_{\varphi_2}$. This leads to a family of conformal K\"ahler structures
on $\C^2$, where the explicit formulae for the conformal factor $y$ and the metric function $w(y,z,\bar{z})$ may be found in 
\cite{Farquet:2014kma}.  In particular one calculates the \emph{local} gauge field given by (\ref{Agauge}) to be
\bea\label{AEAdS}
A_{\mathrm{local}} &=& \frac{\left(b_1+b_2\sqrt{q^2+1}\right)\dd\varphi_1+\left(b_2+b_1\sqrt{q^2+1}\right)
\dd\varphi_2}{2\sqrt{(b_2+b_1\sqrt{q^2+1})^2\cos^2\ppsi +(b_1+b_2\sqrt{q^2+1})^2\sin^2\ppsi }}~,
\eea
which is a non-trivial instanton on Euclidean AdS$_4$. In fact this solution was first found in \cite{Martelli:2011fu} using 
very different methods. One can check that the field strength $F=\dd A$ for (\ref{AEAdS}) indeed defines a smooth, non-singular 
instanton on EAdS$_4$ precisely when $b_1/b_2>0$ or $b_1/b_2=-1$, with $b_1/b_2=\pm 1$ both giving trivial instantons.
When $b_1/b_2<0$ and $b_1/b_2\neq -1$ the instanton is singular along one axis or the other. 

Writing $A$ as a global one-form and restricting to the conformal boundary at $q=\infty$ we obtain
\bea
A_{(0)} &=& \frac{b_2\dd\varphi_1+b_1\dd\varphi_2}{2\sqrt{b_1^2\cos^2\ppsi +b_2^2\sin^2\ppsi }}-\frac{1}{2}(\sign(b_2)\dd\varphi_1+\sign(b_1)\dd\varphi_2)~.
\eea
In particular notice this is well-defined at both poles $\ppsi=0$ and $\ppsi=\pi/2$. 
 The submanifold $\Sigma_2$ is parametrized by the radial direction $q$ in AdS$_4$ and the $S^1$ wrapping $\varphi_1$ or $\varphi_2$ when $\ppsi=0$ or $\ppsi=\pi/2$, respectively.

We now turn to computing (\ref{defSb}). Notice that  the dependence on $b_1$ and $b_2$ arises only via  the gauge field $A$, and not from the metric. Indeed, we compute
\bea
\left[-\int_{\Sigma_2}\vol_{\Sigma_2}+\int_{\de\Sigma_2} \vol_{\de\Sigma_2}\right] & =& 2\pi~,
\eea
and
\bea
\int_{\de\Sigma_2}A_{(0)} & =& \begin{cases}\ \pi(\frac{b_2}{|b_1|}-\sign(b_2))\cdot\sign(b_1) & \quad \mbox{if } \ppsi = 0~, \\ 
\ \pi(\frac{b_1}{|b_2|}-\sign(b_1))\cdot\sign(b_2) & \quad  \mbox{if } \ppsi = \pi/2~. \end{cases}
\eea
The overall factors of $\sign(b_{1})$,  $\sign(b_{2})$ for $\ppsi=0,\pi/2$ arise because the orientation of $\de\Sigma_2$ is determined by $K$. 
Equation \eqref{defSb} immediately gives for all regular cases that
\bea
\mfS_{b_1,b_2} & =& \begin{cases} \displaystyle \ \frac{|b_1|+|b_2|}{2|b_1|} & \quad \mbox{if } \ppsi = 0~, \\ 
\displaystyle \ \frac{|b_1|+|b_2|}{2|b_2|} & \quad \mbox{if } \ppsi = \pi/2~. \end{cases}
\eea
In particular using the variable $\ell$ introduced previously, which is given by $\ell=1/|b_1|$ and $1/|b_2|$ for the $\ppsi=0$ pole and $\ppsi=\pi/2$ pole, respectively, we obtain for both poles and all regular cases that
\bea
\mfS_{b_1,b_2} & =& \frac{|b_1|+|b_2|}{2}\ell~,
\eea
as expected. 

\subsubsection*{Taub-NUT-AdS$_4$}

The Taub-NUT-AdS$_4$ metrics are a one-parameter family of self-dual Einstein metrics on
the four-ball, and have been studied in detail in \cite{Martelli:2011fw}, \cite{Martelli:2012sz}.
The metric may be written 
\bea\label{TNAdSmet}
\dd s^2_4 & =& \frac{r^2-s^2}{\Omega(r)}\dd r^2 + (r^2-s^2)(\ttau^2_1+\ttau_2^2) + \frac{4s^2\Omega(r)}{r^2-s^2}\ttau_3^2~,
\eea
where 
\bea
\Omega(r)  &\equiv  & (r\mp s)^2[1+(r\mp s)(r\pm3s)]~,
\eea
and $\ttau_1,\ttau_2,\ttau_3$  are left-invariant one-forms on $SU(2) \simeq S^3$. The latter may 
 be written in terms of Euler angle variables as
\bea
\ttau_1+\ii\ttau_2 & =& \ex^{-\ii\anglepsi}(\dd\theta+\ii\sin{\theta}\dd\varphi)\ ,\qquad \ttau_3 \ = \ \dd\anglepsi+\cos{\theta}\dd\varphi\ .
\eea
Here $\anglepsi$ has period $4\pi$, while $\theta\in[0,\pi]$ with $\varphi$ having period $2\pi$. The radial coordinate $r$ 
lies in the range $r\in [s,\infty)$, with the origin of the ball $\cong\R^4$ being at $r=s$. 
 The parameter $s>0$ is referred to as the \emph{squashing parameter}, 
with $s=\frac{1}{2}$ being the Euclidean AdS$_4$ metric studied in the previous section. 
The metric is asymptotically locally Euclidean AdS as $r\to\infty$, with
\bea
\dd s^2_4 & \simeq & \frac{\dd r^2}{r^2}+r^2(\ttau_1^2+\ttau_2^2+4 s^2\ttau_3^2)\ ,
\eea
so that the conformal boundary at $r=\infty$ is a biaxially squashed $S^3$.

While the Taub-NUT-AdS metric (\ref{TNAdSmet}) has $SU(2)\times U(1)$ isometry, a 
 generic choice of the Killing vector $K=b_1\partial_{\varphi_1}+b_2\partial_{\varphi_2}=(b_1+b_2)\partial_\varphi+(b_1-b_2)\partial_\anglepsi$ breaks the symmetry of the full solution to $U(1)\times U(1)$. In particular, this symmetry is  broken by the corresponding instanton $A$.
On the other hand, in \cite{Martelli:2011fw}, \cite{Martelli:2012sz} the $SU(2)\times U(1)$ symmetry of the metric was \emph{also} imposed 
on the gauge field, which results in two subfamilies of the above solutions, which are $1/4$ BPS and $1/2$ BPS, respectively. In each case this effectively fixes the  Killing 
vector $K$ (or rather the parameter $b_1/b_2$) as a function of the squashing parameter $s$.

\paragraph{1/4 BPS solution:}The supersymmetric Killing vector for this solution is $K=-\frac{1}{2s}\de_{\anglepsi}$ and we have
\bea
b_1 & = & -b_2  \ = \ -\frac{1}{4s}\ .
\eea
Here $\anglepsi=\varphi_1-\varphi_2$, $\varphi=\varphi_1+\varphi_2$ is the change of angular 
coordinates.
The boundary gauge field $A_{(0)}$ is \cite{Martelli:2012sz}
\bea
A _{(0)}&=& \frac{1}{2}(4s^2-1)\ttau_3 \ ,
\eea
which is a global one-form on $M_3\cong S^3$. 
We may now take the surface $\Sigma_2$ wrapped by the M2-brane to be \emph{any} $S^1$ orbit 
of the Hopf Killing vector $\partial_\anglepsi$ (at any point on the base $S^2=S^3/U(1)_{\anglepsi}$), together with the radial direction $r$. 
This is supersymmetric, and the regularized   volume of $\Sigma_2$ is
\bea
\left[-\int_{\Sigma_2}\vol_{\Sigma_2}+\int_{\de\Sigma_2} \vol_{\de\Sigma_2}\right] & =& 8\pi s^2\ ,
\eea
while the gauge field integral is
\bea
\int_{\de\Sigma_2}A_{(0)} & =& -2\pi (4s^2-1)\ .
\eea
This leads to
\bea
\mfS_{b_1,b_2} & =& 1 \ = \ \frac{|b_1|+|b_2|}{2}\ell\ ,
\eea
where $\ell=4s$ is the length of $K$ divided by $2\pi$. 

\paragraph{1/2 BPS solution:}

The Taub-NUT-AdS metric  (\ref{TNAdSmet}) also admits a 1/2 BPS solution 
\cite{Martelli:2011fw}, \cite{ Martelli:2012sz}. There are thus two linearly independent Killing spinors, 
and an appropriate linear combination preserves $U(1)\times U(1)$ symmetry, 
leading to the Killing vector
\bea
K & =& \left(2 s + \sqrt{4 s^2 - 1}\right)\de_{\varphi}+\left(\tfrac{1}{2 s} - 2 s - \sqrt{4 s^2 - 1}\right)\de_{\anglepsi}~,
\eea
so that
\bea
b_1 & =& \frac{1}{4s}\ ,\qquad b_2\ = \ -\frac{1}{4s}+2s+\sqrt{4s^2-1}\ .
\eea
The boundary gauge field is
\bea
A_{(0)} &=& s\sqrt{4s^2-1}\ttau_3\ .
\eea
This time we take the Wilson loop to wrap one of the two poles $\theta=0$, $\theta=\pi$. 
These are both copies of $S^1$, and $\Sigma_2$ is again formed by adding the 
radial direction $r$.  The boundary gauge field is
\bea
A_{(0)}\mid_{\mathrm{pole}} & =& \begin{cases}\  \ \ 2s\sqrt{4s^2-1}\ \dd\varphi_1 & \quad \mbox{if } \theta = 0~, \\ \  
-2s\sqrt{4s^2-1}\ \dd\varphi_2 & \quad \mbox{if } \theta = \pi~. \end{cases}
\eea
The regularized volume is again $8\pi s^2$, which then gives
\bea
\mfS_{b_1,b_2} & =& \begin{cases} \ 2s(2s+\sqrt{4s^2-1}) & \quad \mbox{if } \theta = 0~, \\ 
\ 2s(2s-\sqrt{4s^2-1}) & \quad  \mbox{if } \theta = \pi~. \end{cases}
\eea
In both cases we indeed have
\bea
\mfS_{b_1,b_2} & =& \frac{|b_1|+|b_2|}{2}\ell\ ,
\eea
where $\ell=1/|b_{1}|$, $\ell=1/|b_{2}|$ for the two poles.

\section{Discussion}\label{discuss}

In this paper we have derived the formula 
(\ref{Wilvevres}), (\ref{defSb1b2}) for the expectation values of large $N$ BPS Wilson loops, in both 
gauge theory and in supergravity. In particular the gauge theories are defined on a general class 
of supersymmetric backgrounds $M_3\cong S^3$, which in the supergravity dual arise 
as the conformal boundaries of self-dual solutions to gauged supergravity. A key feature 
of the gravity calculation is that we are able to evaluate the regularized M2-brane action, 
that is identified with the Wilson loop VEV, without using the explicit form of the metric and graviphoton field. 
This seems to be a general feature of such computations of BPS quantities in AdS/CFT, and allows 
us to verify the correspondence for these observables in a very broad class of solutions.

The results described in this paper lead to a number of questions, and possible future directions to pursue. 
First, in supergravity we have restricted to self-dual solutions, while more generally there 
are also non-self-dual solutions to gauged supergravity. A local study of these solutions 
appears in \cite{Dunajski:2010uv}, while global asymptotically locally Euclidean AdS solutions were constructed in \cite{Martelli:2012sz}.
Presumably the methods we have used extend to this general class of solutions. In particular 
the Wilson loop was computed for a charged topological black hole background in \cite{Nishioka:2014mwa}, 
and successfully compared to a field theory calculation.
The non-self-dual solutions 
in \cite{Martelli:2012sz} all have the feature that the bulk $M_4$ has non-trivial topology, which in turn 
leads to issues in interpreting them holographically (and in particular uplifting to eleven dimensions 
restricts the choice of $Y_7$, implying the solutions are only relevant to specific gauge theories on $M_3$). It would be interesting to try to calculate Wilson loops 
in such examples, and compare to a dual field theory computation. Indeed, in \cite{Martelli:2012sz}
it was argued that in appropriate circumstances $\langle \, W\, \rangle =0$ in supergravity, via 
a similar mechanism to that in \cite{Witten:1998zw}. 
Typically here the boundary $M_3$ in such examples has a 
non-trivial fundamental group, as in the large $N$ gauge theory computation in \cite{Alday:2012au}, 
and there is indeed evidence that if the R-symmetry gauge field on $M_3$ has non-trivial 
topology then the large $N$ Wilson loop VEV vanishes also in the gauge theory.\footnote{The matrix model 
behaviour is then much more complicated, and seems difficult to analyse analytically, but very
roughly speaking the Wilson loop VEV averages to zero due to the sum $\sum_{k=0}^{p-1}\omega_p^k=0$, 
where $\omega_p$ is a primitive $p$th root of unity. This arises from the fact that the dominant contribution 
to the Wilson loop at large $N$ comes from a non-trivial flat connection, with $\omega_p^k$ related to its holonomies 
\cite{AMS}. } 
Finally, it is now clear that similar results should also hold in higher dimensions. A very similar formula 
to (\ref{Wilvevres}), (\ref{defSb1b2}) was found to hold for certain supersymmetric squashed 
five-sphere conformal boundaries and their gravity duals in \cite{Alday:2014rxa}, \cite{Alday:2014bta}, and 
was conjectured to hold for general backgrounds in those references. It would also be interesting to compute 
Wilson loops in the general class of $S^1\times S^3$ Hopf surface geometries in \cite{Assel:2014paa}.

\subsection*{Acknowledgments}
\noindent
We would like to thank D.~Martelli for comments. D.~F. is supported by the Berrow Foundation and J.~F.~S. is supported by the Royal Society.

\end{document}